\documentclass[aip,graphicx]{revtex4-1}

\usepackage{graphicx}
\usepackage{dcolumn}
\usepackage{bm}

\usepackage{subfigure}
\usepackage{float}
\newsavebox{\measurebox}
\usepackage{amsfonts}
\usepackage{amsmath}

\usepackage[utf8]{inputenc}
\usepackage[T1]{fontenc}
\usepackage{mathptmx}
\usepackage{etoolbox}

\draft 

\begin{document}


\title{An application of data driven reward of deep reinforcement learning by dynamic mode decomposition in active flow control
} 



\author{Sheng Qin}
    \affiliation{\quad Department of Aeronautics and Astronautics, Fudan University}
\author{Shuyue Wang}
    \affiliation{\quad Department of Aeronautics and Astronautics, Fudan University}
\author{Jean Rabault}
    \affiliation{\quad Information Technology Department, Norwegian Meteorological Institute}
\author{Gang Sun}
    \email{Email address for correspondence: gang\_sun@fudan.edu.cn}
    \affiliation{\quad Department of Aeronautics and Astronautics, Fudan University}


\date{\today}

\begin{abstract}
\textbf{ABSTRACT:}
This paper focuses on the active flow control (AFC) of the flow over a circular cylinder with synthetic jets through deep reinforcement learning (DRL) by  implementing a reward function based on dynamic mode decomposition (DMD).
As a main factor that affects the DRL model, the reward is determined by the information extracted from flow field by performing DMD on measurements through simulation.
With the data-driven reward, the DRL model is able to learn the AFC policy through the more global information of the field, and instructs the mass flow rate of the synthetic jets.
As a result of this type of AFC, the vortex street is stabilized with a reduction of approximately $8\%$ in drag and an improvement of approximately $109\%$ in recirculation area.
Furthermore, the configuration of the flow modified by the AFC is studied with DMD on the velocity measurement of the complete flow field.
\end{abstract}


\maketitle 

\section{INTRODUCTION}
K\'arm\'an vortex street is a famous phenomenon when flow separating from a bluff body such as a circular cylinder~\cite{roshko1955wake, berger1972periodic, williamson2004vortex}. 
The periodic flow separation and instability cause vibrations and noise, which can be weakened through passive~\cite{bearman1965investigation, bearman1998reduction, bagheri2012spontaneous, deng2019dynamics} and active~\cite{poncet2004topological, amitay1998aerodynamic, mittal2001control} methods.
Active flow control (AFC) has been studied by a lot of researchers and has witnessed a fast growth~\cite{choi2008control}.
Various forcing devices have been applied on the active control of flow over a bluff body, such as rotary~\cite{poncet2004topological}, streamwise~\cite{cetiner2001streamwise, leontini2013wake}, and transverse\cite{blackburn1999study, kumar2016experimental} oscillations of the bluff body, distributed forcing~\cite{kim2005distributed}, synthetic jets~\cite{amitay1998aerodynamic, smith1998formation, glezer2002synthetic}, rotating control cylinders~\cite{mittal2001control, zhu2015simultaneous, schulmeister2017flow}, etc.

Besides the various control approaches, the control policies of AFC are worth further investigation. 
The high non-linearity of the dynamic system of AFC makes it challenge to find efficient control policies.
For the studies of open-loop AFC (no feedback mechanism), a periodic signal is often used as the control policy in the AFC of flow over a bluff body~\cite{nakano1994flow, konstantinidis2005timing, jeon2004active, fujisawa2004phase}.
In these cases, the control system cannot interact with the feedback of flow field.
For the active closed-loop control cases,  many  control algorithms based on mathematical analysis are utilized in the determination of the control policies of AFC, such as optimal control theory~\cite{min1999suboptimal, he2000active, protas2002optimal, li2003optimal} and reduced-order models~\cite{cortelezzi1997nonlinear, gillies1998low, li2003feedback, protas2004linear, RN7}.
However, these model-based control policies are usually based on either harmonic or constant forcing~\cite{brunton2015closed, schoppa1998large}.
The complex high-dimensional nonlinear systems of actual AFC are challenge to these conventional control algorithms.

In recent years, many artificial intelligence techniques caught the attention of researchers~\cite{brunton2020machine}. 
Reinforcement learning (RL) is one such approach that interacts with the environment according to policies, learns from experience, and improves policies. 
In the case of RL, an agent (controlled by the artificial neural networks (ANNs) described in Section~\ref{sec4}) interacts with an environment through three channels of exchange of information in a closed-loop fashion~\cite{sutton2018reinforcement}. 
First, the agent is given access at each time step to a state $s_t$ of the environment. 
Second, the agent performs an action $a_t$, that influences the time evolution of the environment.
Finally, the agent receives a reward $r_t$ depending on the state of the environment following the action.
The selection of reward function is important and related to the learning result.
The RL model keep on finding policies learnt from the experience of interacting with the environment, in order to discover control sequences that yield the highest possible reward.
The environment can be any system that can provides the interface $(s_t,a_t,r_t)$.
As a model-free approach, RL is quite suitable for complex, high-dimensional, nonlinear systems.
With deep ANNs used in the framework, deep reinforcement learning (DRL) can extract features from data in high-dimensional state and action spaces.
DRL has been successfully applied to resolve several problems such as: playing Atari games~\cite{mnih2013playing}, controlling complex robots~\cite{kober2013reinforcement, gu2017deep} and generating realistic dialogs~\cite{li2016deep}.

Since these challenging systems successfully controlled by DRL have the properties of high-dimension and non-linearity similar to the features of AFC, the DRL arouses researchers' interests as one new approach of AFC.
In recent years, DRL has been applied on AFC in several cases.
Rabault et al.~\cite{RN3} introduced DRL technique to the AFC problem of flow over a circular cylinder.
They used synthetic jets set on the surface of the cylinder, injecting and suctioning mass flow into the field to control the flow.
With the control policies learnt through DRL, the mass flow rate of the jets is controlled to stabilize the vortex alley in a short period of time.
Based on this framework, Xu et al.~\cite{RN1} placed two small rotational cylinders behind the main cylinder to control the flow.
The rotation law of the small cylinders is directed by DRL model.
Xu et al. also achieved a good result on stabilizing the vortex alley, which shows that the DRL-based AFC method performs well on the flow over a cylinder through various control approaches like synthetic jets and rotational cylinders.
Tang et al.~\cite{RN2} analyzed the AFC based on DRL at different Reynolds number conditions. 
The results demonstrated that this AFC method through DRL technique is robust and has good controlling performance over a range of Reynolds number.

In Rabault et al.'s work, the reward of the DRL model is based on the drag coefficient and lift coefficient of the cylinder~\cite{RN3}. 
The drag coefficient and lift coefficient are extracted from flow field, where a lot of other detailed information are ignored or not considered in computing the coefficients.
Compared with the total performance, the complete flow field snapshots contains the whole information for analysis and control of the flow, which can lead to a finer result.
However, the extremely high-dimensional data of the flow field snapshots is highly memory-cost and unfeasible for computation. 
Seeking a low-dimensional description from the snapshots while keeping dynamically important flow field information is necessary and important.
One way to achieve this goal is using a reduced order model (ROM).
The basic idea of ROM is to obtain a low-dimensional description of the flow features which provides the spatiotemporal information of a flow field needed for feedback control through some matrix analysis means such as proper orthogonal decomposition (POD) and dynamic mode decomposition (DMD)~\cite{kutz2016dynamic}.

With the application of ROM, the dynamically important information of the periodic vortex shedding can be extracted from the complete flow field information, and be used to make AFC policy.
As a mainstream ROM, POD~\cite{RN14} has been widely used on the AFC of the flow over a circular cylinder based on various 
forcing devices, such as rotary~\cite{RN6, RN7}, transverse displacement~\cite{RN8} and synthetic jets~\cite{RN15, RN20}.
Another popular ROM which has been applied in flow field analysis and AFC in recent years is DMD.
Dynamic mode decomposition, proposed by Schmid~\cite{RN10}, extracts the spatiotemporal information of the flow and constructs a low-dimensional, linearized ROM for the high-dimensional flow field snapshots and flow evolution.
Different from POD which computes orthogonal modes based on the energy information, DMD can capture dominant flow modes with purely frequency content~\cite{kou2019dynamic}. 
Although having extensive applications on flow field analysis, the standard DMD mainly focus on the basic periodic dynamics of the flow, which has limitations for problems with external actuation or external control.
As a developed method from DMD, dynamic mode decomposition with control (DMDc)~\cite{RN11}, can consider the effect of control and extract more accurate reduced-order models and dynamics modes for the complex systems with external control.
As a good approach to catch modes of the flow, DMD has been utilized to analyse the computational and experimental data of the flow over a bluff body in many studies~\cite{tu2013dynamic, hemati2014dynamic, bagheri2013koopman, he2013initial, zhang2014identification}.

For its application on AFC, a study on bistable flow of the flow around a circular cylinder under the synthetic-jet control was carried out by Wang et al.~\cite{wang2017extraction} through total DMD (TDMD).
DMD was employed to extract dynamic modes, spectrum, and mode coefficients to describe the flow dynamics to reveal the underlying flow physics of the bistable flow.
The result showed that the bistable flow contains dual dominant frequencies corresponding to the antisymmetric and symmetric vortex-shedding modes and the DMD method provides an effective and efficient approach to analyze the bistable flow.
Jardin and Bury~\cite{jardin2012lagrangian} investigated the influence of the control of pulsed tangential jets on the flow past a circular cylinder.
The time-series data were analysed using a spectral–Lagrangian dual
approach based on DMD and vortex tracking, which revealed the interaction mechanisms (between the tangential jets and the separated shear layers) governing the forced flow.
To control the flow over a circular cylinder, Shaafi and Vengadesan~\cite{shaafi2014wall} set a control rod of the same diameter upstream the cylinder.
The control rod was subjected to different rotation rates. 
Modal structures obtained from DMD revealed that the flow structures behind the main cylinder are suppressed due to wall and the flow is dominated by the wake of control rod.

In this paper, the AFC through DRL applied on the flow over a circular cylinder is studied.
To achieve a reward function that contains more global flow field information, DMDc is performed on the snapshots to extract the temporal and spatial characteristics of the flow field.
The modes and the corresponding frequencies of the flow with AFC can be computed via DMDc.
To stabilize the vortex street
, the amplitudes of each dynamic modes are expected to decrease.
With the decomposition of the flow field information, a data-driven reward function of the DRL network is computed from the combination of these mode amplitudes.
With the improved reward function, the modified AFC method based on DRL is performed on the flow over a circular cylinder and achieve good control results.
With the active control guided by the policy learnt from the flow field information, the drag of the flow over a circular cylinder is reduced, as well as the fluctuations of drag and lift.
In addition, the recirculation area (the downstream region of the cylinder where the horizontal component of the velocity is negative) of the controlled flow is increased and the low-velocity region in the wake is extended.
Furthermore, a DMD analysis is conducted on the complete field of the baseline flow and the flow with control for detailed information and the comparison of the dynamic modes of the baseline flow and the controlled flow is discussed. 

This paper is organized as follows: In Section~\ref{sec2}, the simulation base of the flow over a circular cylinder controlled by synthetic jets is introduced.
In Section~\ref{sec3}, an introduction to the DMD and DMDc algorithms is presented.
Section~\ref{sec4} introduces DRL framework and its application, and proposes a new data-driven reward for the learning in this paper.
Section~\ref{sec5} shows the controlling results and the DMD analysis of it, and the paper is concluded in Section~\ref{sec6}.

\section{SIMULATION ENVIRONMENT}\label{sec2}
\begin{figure}[H]
\includegraphics[width=1.0\textwidth]{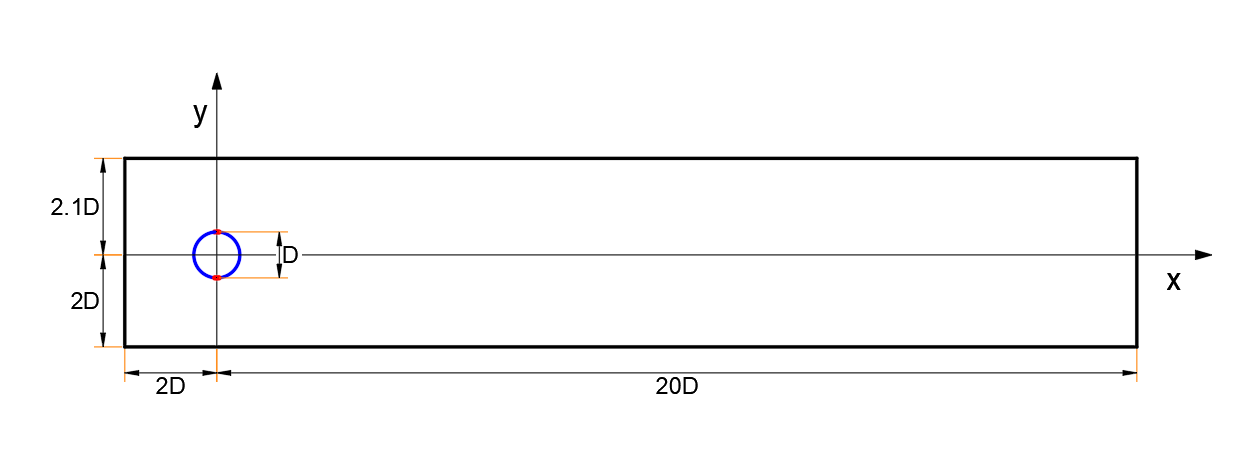}
\caption{\label{fig:geometry} Geometrical description of the configuration used for the simulation adapted from Sch\"afer et al.'s work. 
The synthentic jets are marked with red arcs.}
\end{figure}

The synthetic-jet AFC is applied on the flow past a circular cylinder immersed in a two-dimensional channel, with the configuration of the simulation adapted from the classical benchmark computations carried out by Sch\"afer~\cite{RN17}.
The simulation acts as the environment that the DRL agent interacts with.
With all quantities considered non-dimensionalized, the geometry of the simulation is shown in Fig.~\ref{fig:geometry}.
The benchmark consists of a circular cylinder of non-dimensional diameter $D=1$ placed in a two-dimensional domain of total non-dimensional length $L=22$ and height $H=4.1$.
The center of the cylinder is located at a shift of 0.05 from the horizontal centreline of the flow domain.
Two jets normal to the cylinder wall are symmetrically implemented on the upper and lower sides of the cylinder, at angles $\theta_1=90^{\circ}$ and  $\theta_2=270^{\circ}$ relative to the ﬂow direction.
The jets are controlled through their non-dimensional mass flow rates, $Q_i,~(i=1,2)$.
The total mass flow rate injected by the jets is set to zero, i.e. $Q_1+Q_2=0$.
Therefore, the AFC in this case is in fact according to one scalar value $Q_1$.
Negative values of $Q_i$ correspond to suction.

In this viscous, incompressive flow, the governing equations are the two-dimensional, time-dependent Navier–Stokes equations and the continuity equation, expressed in a non-dimensional form:\begin{equation}
    \frac{\partial\boldsymbol{u}}{\partial{t}}+\boldsymbol{u}\cdot(\nabla\boldsymbol{u})=-\nabla{p}+\frac{1}{Re}\Delta\boldsymbol{u}
\end{equation}
\begin{equation}
    \nabla\cdot\boldsymbol{u}=0
\end{equation}
The Reynolds number is defined as $Re=\bar{U}D/\nu$, where $\nu$ is the  kinematic viscosity and the characteristic velocity $\bar{U}$ is the mean inflow velocity magnitude.

The inflow velocity profile, according to Sch\"afer~\cite{RN17}, can be expressed as follows: \begin{equation}
    U(y)=6(H/2-y)(H/2+y)/H^2
\end{equation}
where $U(y)$ is the horizontal non-dimensionalized velocity component\underline{ at the inlet}.
The mean velocity magnitude is $\bar{U}=2U(0)/3=1$.
No-slip boundary conditions are imposed on the top and bottom walls and on the solid walls of the cylinder.
An outﬂow boundary condition is imposed based on the assumption that the derivative of the velocity along the X-axis is zero at the outlet.
The Reynolds number in this case is set to $Re=100$.
For the boundary conditions of synthetic jets, the radial velocity profiles of the two jets are set as: \begin{equation}
    u_{jet}(\theta,Q_i)=\frac{\pi}{\omega{}D}Q_i\text{cos}\left(\frac{\pi}{\omega}(\theta-\theta_i)\right)
\end{equation}
where $\theta$ is the angular coordinate and $\theta_i,~(i=1,2)$ is the angular position of the center of each jet.
$\omega=10^{\circ}$ is the width of each jet.

The computational mesh is generated by Gmsh~\cite{RN19}, an open-source finite element mesh generator.
The computational domain is discretized into triangular cells, and refined around the cylinder surface.
Based on a finite-element framework FEniCS~\cite{LoggMardalEtAl2012a, AlnaesBlechta2015a}, the governing Navier-Stokes equations are solved in a segregated manner~\cite{valen2012comparison}.
More precisely, the incremental pressure correction scheme (IPCS) method~\cite{RN18} with an explicit treatment of the nonlinear term is used.

A non-dimensional, constant numerical time step $dt=5\times10^{-3}$ is used.
The drag and lift forces on the cylinder at each time step are calculated respectively as \begin{equation}
    F_D=\int_C(\boldsymbol{\sigma}\cdot\boldsymbol{n})\cdot\boldsymbol{e}_xdS
\end{equation}
and \begin{equation}
    F_L=\int_C(\boldsymbol{\sigma}\cdot\boldsymbol{n})\cdot\boldsymbol{e}_ydS
\end{equation}
where $\boldsymbol{\sigma}$ is the Cauchy stress tensor, $\boldsymbol{n}$ is the unit vector normal to the cylinder surface, and $\boldsymbol{e}_x=(1,0)$ and $\boldsymbol{e}_y=(0,1)$.
They are normalized into drag coefﬁcient and lift coefﬁcient \begin{equation}
    C_D=\frac{F_D}{\rho\bar{U}^2D/2}
\end{equation}
and \begin{equation}
    C_L=\frac{F_L}{\rho\bar{U}^2D/2}
\end{equation}
The Strouhal number (St) is defined as $St=fD/\bar{U}$, where $f$ is the vortex shedding frequency computed by performing fast Fourier transform (FFT) on the lift coefficients $C_L$ depending on time.
The injected mass flow rates are normalized as $Q_i^*=Q_i/Q_{ref}$, where $Q_{ref}=\int_{-D/2}^{D/2}\rho{}U(y)dy$.
The normalized mass flow rates $Q_i^*$ is constrained in this study as $\left|Q_i^*\right|<0.068$.

\begin{figure}[H]
\includegraphics[width=1.0\textwidth]{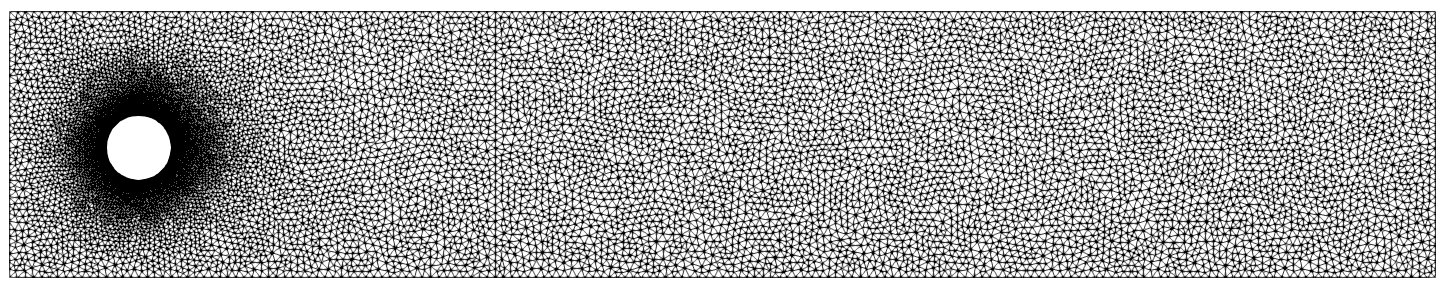}
\caption{\label{fig:mesh} Two-dimensional unstructured mesh of the computation domain of the flow over a circular cylinder.}
\end{figure}

The mesh-independence study for this benchmark is conducted, with meshes of three different resolutions, which is shown in Table~\ref{tab:mesh}.
The maximum value of $C_D$ in this case is within $0.2\%$ of what is reported in the benchmark of Sch\"afer et al.~\cite{RN17}, and similar agreement is found for $St$,  which validates the simulation.
As can be seen in TABLE~\ref{tab:mesh},the resolution of the main mesh, which is used in this paper, is fine enough for the simulation with the discrepancies less than $0.04\%$ in all listed quantities when compared with the fine mesh.
The mesh used for the simulation is plotted in Fig.~\ref{fig:mesh}.

\begin{table}
\caption{\label{tab:mesh}Flow parameters for the flow over a circular cylinder of different mesh resolution. (The numbers of mesh cells are listed in the first column.)}
\begin{ruledtabular}
\begin{tabular}{lccc}
Mesh resolution & $C_D^{max}$ & $C_D^{max}$ & $St$\\
\hline
Coarse 9252 & 1.075 & 3.242 & 0.3027\\
Main 25002  & 1.077 & 3.245 & 0.3027\\
Fine 104345 & 1.077 & 3.246 & 0.3027\\
\end{tabular}
\end{ruledtabular}
\end{table}

\section{DYNAMIC MODE DECOMPOSITION}\label{sec3}
\subsection{Basic dynamic mode decomposition (DMD)}
An brief introduction to the standard DMD algorithm is present in this section.
The DMD algorithm starts with representing the data of flow field into vectors in the form of sequential snapshots, with constant time step $\Delta{t}$.
The vectors of $N$ snapshots can be arranged into a matrix $\boldsymbol{V}_1^N$:
\begin{equation}
    \boldsymbol{V}_1^N=\{\boldsymbol{v}_1,\boldsymbol{v}_2,\boldsymbol{v}_3,...,\boldsymbol{v}_N\},
\end{equation}
where $\boldsymbol{v}_i$ stands for the $i$th snapshot of flow field. 

Assuming that the flow field $\boldsymbol{v}_i$ can be connected to the subsequent flow field $\boldsymbol{v}_{i+1}$ with a linear mapping, adjacent snapshot vectors can be approximately expressed by:
\begin{equation}
    \boldsymbol{v}_{i+1}=\boldsymbol{Av}_i,
\end{equation}
where $\boldsymbol{A}$ is the system matrix.
This approximation is assumed to hold for all adjacent snapshots. Then the subsequent snapshot matrix $\boldsymbol{X_1}=\boldsymbol{V}_2^m$ can be represented by the current snapshot matrix $\boldsymbol{X_0}=\boldsymbol{V}_1^{m-1}$ and the system matrix $\boldsymbol{A}$, that is,
\begin{equation}
    \boldsymbol{X_1}=\boldsymbol{AX}_0.
    \label{eq:system}
\end{equation}

The primary objective of DMD is to solve for an approximation of the high-dimensional system matrix $\boldsymbol{A}$ for the snapshot matrix pair $\boldsymbol{X_0}$ and $\boldsymbol{X_1}$.
A least-squares solution $\boldsymbol{A}$ to the underdetermined problem described in Eq.~\ref{eq:system} can be achieved by minimizing the Frobenius norm of $\Vert\boldsymbol{X}_1-\boldsymbol{AX}_0\Vert_F$.
One computationally efficient and accurate way to find the dynamic modes and eigenvalues of the underlying system matrix $\boldsymbol{A}$ is solving a similar matrix $\boldsymbol{\tilde{A}}$ instead of the full-order $\boldsymbol{A}$, via similarity transformation.

To seek an invertible matrix for the similarity transformation, the singular value decomposition (SVD) of the snapshot matrix $\boldsymbol{X}_0$ is computed,
\begin{equation}
    \boldsymbol{X}_0=\boldsymbol{U\Sigma{}V}^*,
\end{equation}
where $\boldsymbol{U}^*\boldsymbol{U}=\boldsymbol{I}$ and $\boldsymbol{V}^H\boldsymbol{V}=\boldsymbol{I}$. 
$^*$ denotes the complex conjugate transpose.
$\boldsymbol{\Sigma}$ is a diagonal matrix containing the singular values. The system matrix $\boldsymbol{A}$ can be transformed into the similar matrix $\boldsymbol{\tilde{A}}$:
\begin{equation}
    \boldsymbol{A}=\boldsymbol{U\tilde{A}U}^*.
\end{equation}
With the similarity transformation, the problem of minimizing $\Vert\boldsymbol{X}_1-\boldsymbol{AX}_0\Vert_F$ turns into minimizing the Frobenius norm of $\Vert\boldsymbol{X}_1-\boldsymbol{U\tilde{A}\Sigma{}V}^*\Vert_F$.
Then the approximation of $\boldsymbol{A}$ can be solved:
\begin{equation}
    \boldsymbol{A}\approx\boldsymbol{\tilde{A}}=\boldsymbol{U}^*\boldsymbol{X}_1\boldsymbol{V\Sigma}^{-1}.
    \label{eq:A_tilde}
\end{equation}
The matrix $\boldsymbol{\tilde{A}}$ defines a low-dimensional linear dynamical system: 
\begin{equation}
    \boldsymbol{\tilde{v}}_{i+1}=\boldsymbol{\tilde{A}\tilde{v}}_i,
\end{equation}
where $\boldsymbol{\tilde{v}}_i$ can be used to reconstruct the high-dimensional flow field $\boldsymbol{v}_i=\boldsymbol{U\tilde{v}}_i$.

The eigendecomposition of is first performed on $\boldsymbol{\tilde{A}}$:
\begin{equation}
    \boldsymbol{\tilde{A}W}=\boldsymbol{W\Lambda},
\end{equation}
where $\boldsymbol{\Lambda}$ and $\boldsymbol{W}$ consist of eigenvalues and eigenvectors of $\boldsymbol{\tilde{A}}$, respectively.
Because $\boldsymbol{\tilde{A}}$ is the similar matrix of $\boldsymbol{A}$, they have the same eigenvalues, which means that $\boldsymbol{\Lambda}$ gives the eigenvalues of $\boldsymbol{A}$. 
Reconstructed from $\boldsymbol{W}$, the eigenvectors of $\boldsymbol{A}$ (DMD modes) are given by columns of $\boldsymbol{\Phi}$ (referring to Eq.~\ref{eq:A_tilde}):
\begin{equation}
    \boldsymbol{\Phi}=\boldsymbol{X}_1\boldsymbol{V\Sigma}^{-1}\boldsymbol{W}.
\end{equation}

For the $i$th DMD mode $\boldsymbol{\phi}_i$ and the corresponding eigenvalue $\mu_i$, the growth $g_i$ rate and physical frequency $\omega_i$ of this mode are:
\begin{equation}
    g_i=\text{Re}[\text{log}(\mu_j)]/\Delta{}t,
\end{equation}  
\begin{equation}
    \omega_i=\text{Im}[\text{log}(\mu_j)]/\Delta{}t.
\end{equation}

\subsection{Dynamic mode decomposition with control (DMDc)}
As a developed method from DMD, dynamic mode decomposition with control (DMDc), proposed by Proctor et al.~\cite{RN11}, can consider the effect of control and extract more accurate reduced-order models for the complex systems with control.
The DMDc method modifies the assumption on linear system of DMD, by comprising the control snapshot in the system equation: \begin{equation}
    \boldsymbol{v}_{i+1}=\boldsymbol{Av}_i+\boldsymbol{Bu}_i,
\end{equation}
where $\boldsymbol{v}_{i},\boldsymbol{v}_{i+1}\in\mathbb{R}^n$, $\boldsymbol{u}_{i}\in\mathbb{R}^l$, $\boldsymbol{A}\in\mathbb{R}^{n\times{}n}$ and $\boldsymbol{B}\in\mathbb{R}^{n\times{}l}$.

The vectors of control snapshots can be arranged into a matrix $\boldsymbol{\Gamma}$: \begin{equation}
    \boldsymbol{\Gamma}=\{\boldsymbol{u}_1,\boldsymbol{u}_2,\boldsymbol{u}_3,...,\boldsymbol{u}_{m-1}\},
\end{equation}
where $\boldsymbol{u}_i$ stands for the $i$th control snapshot. Then Eq.~\ref{eq:system} can be rewritten in matrix form: \begin{equation}
    \boldsymbol{X_1}=\boldsymbol{AX}_0+\boldsymbol{B\Gamma}.\label{eq:snapshots}
\end{equation}

The approximate relationship among the flow flied snapshot matrices $\boldsymbol{X}_0$, $\boldsymbol{X}_1$ and control snapshot matrix $\boldsymbol{\Gamma}$ can be rewritten as follows: \begin{equation}
    \boldsymbol{X}_1=\boldsymbol{G\Omega},
    \label{eq:system2}
\end{equation}
where $\boldsymbol{G}=[\boldsymbol{A},~ \boldsymbol{B}]$ and $\boldsymbol{\Omega}=\left[ \begin{array}{cc}
    \boldsymbol{X}_0 \\ 
    \boldsymbol{\Gamma}
\end{array} \right]$. Similar to DMD, the operator $\boldsymbol{G}$ can be computed as \begin{equation}
    \boldsymbol{G}=\boldsymbol{X}_1\boldsymbol{\Omega}^{\dagger},
\end{equation}
where $\dagger$ refers to the Moore–Penrose pseudoinverse.
The best-fit solution of the operator $\boldsymbol{G}$  to the underdetermined problem of Eq.~\ref{eq:system2} can be found by minimizing the Frobenius norm
of $\Vert\boldsymbol{X}_1-\boldsymbol{G\Omega}\Vert_F$.
By performing SVD on $\boldsymbol{\Omega}$ with truncation value $p$ as $\boldsymbol{\Omega}\approx\boldsymbol{\tilde{U}\tilde{\Sigma}\tilde{V}}^*$, an approximation of $\boldsymbol{G}$ can be computed by the following: \begin{equation}
    \boldsymbol{G}\approx\boldsymbol{\tilde{G}}=\boldsymbol{X}_1\boldsymbol{\tilde{V}\tilde{\Sigma}}^{-1}\boldsymbol{\tilde{U}}^*,
\end{equation}
where $\boldsymbol{G}\in\mathbb{R}^{n\times(n+l)}$.
The approximations of the matrices $\boldsymbol{A}$ and $\boldsymbol{B}$ can be computed by breaking
the linear operator $\boldsymbol{U}$ into two separate components as follows: \begin{equation}
    \begin{aligned}
    \left[\boldsymbol{A},~ \boldsymbol{B}\right]&\approx\left[\boldsymbol{\bar{A}},~ \boldsymbol{\bar{B}}\right] \\
    &=\left[\boldsymbol{X}_1\boldsymbol{\tilde{V}\tilde{\Sigma}}^{-1}\boldsymbol{\tilde{U}}^*_1,~ \boldsymbol{X}_1\boldsymbol{\tilde{V}\tilde{\Sigma}}^{-1}\boldsymbol{\tilde{U}}^*_2\right],
    \end{aligned}
\end{equation}
where $\boldsymbol{\tilde{U}}^*_1\in\mathbb{R}^{n\times{}p}$, $\boldsymbol{\tilde{U}}^*_2\in\mathbb{R}^{l\times{}p}$ and $\boldsymbol{\tilde{U}}^*=\left[\boldsymbol{\tilde{U}}^*_1,~ \boldsymbol{\tilde{U}}^*_2\right]$.

A second SVD is utilized to find the reduced-order subspace of the output space.
The output matrix $\boldsymbol{X_2}$ can be approximated by performing SVD with truncation value $r$ as $\boldsymbol{X_2}\approx\boldsymbol{\hat{U}\hat{\Sigma}\hat{V}}^*$, where $\boldsymbol{\hat{U}}\in\mathbb{R}^{n\times{}r}$, $\boldsymbol{\hat{\Sigma}}\in\mathbb{R}^{r\times{}r}$ and $\boldsymbol{\hat{V}}^*\in\mathbb{R}^{r\times{}(m-1)}$.
Therefore, the approximation of $\boldsymbol{G}$ can be computed by $\boldsymbol{G}=\left[\boldsymbol{A},~\boldsymbol{B}\right]\approx\left[\boldsymbol{\hat{U}\tilde{A}\hat{U}}^*,~\boldsymbol{\hat{U}\tilde{B}}\right]$, and the reduced-order approximations of $\boldsymbol{A}$ and $\boldsymbol{B}$ can be computed as follows: \begin{equation}
    \boldsymbol{\tilde{A}}=\boldsymbol{\hat{U}}^*\boldsymbol{\bar{A}\hat{U}}=\boldsymbol{\hat{U}}^*\boldsymbol{X}_1\boldsymbol{\tilde{V}\tilde{\Sigma}}^{-1}\boldsymbol{\tilde{U}}^*_1\boldsymbol{\hat{U}},
\end{equation}
\begin{equation}
    \boldsymbol{\tilde{B}}=\boldsymbol{\hat{U}}^*\boldsymbol{\bar{B}}=\boldsymbol{\hat{U}}^*\boldsymbol{X}_1\boldsymbol{\tilde{V}\tilde{\Sigma}}^{-1}\boldsymbol{\tilde{U}}^*_2,
\end{equation}
where $\boldsymbol{\tilde{A}}\in\mathbb{R}^{r\times{}r}$ and $\boldsymbol{\tilde{B}}\in\mathbb{R}^{r\times{}l}$.

Similar to DMD, the dynamic modes of $\boldsymbol{A}$ can be found by first solving the eigenvalue decomposition $\boldsymbol{\tilde{A}W}=\boldsymbol{W\Lambda}$. 
Then the dynamic modes of the operator $\boldsymbol{A}$ can be computed as \begin{equation}
    \boldsymbol{\Phi}=\boldsymbol{X}_1\boldsymbol{\tilde{V}\tilde{\Sigma}}^{-1}\boldsymbol{\tilde{U}}^*_1\boldsymbol{\hat{U}}\boldsymbol{W},
\end{equation}
and the corresponding amplitudes of the dynamic modes, $(\alpha_1,\alpha_2,...,\alpha_r)$ can be computed by: \begin{equation}
    (\alpha_1,\alpha_2,...,\alpha_r)^\mathrm{T}=\boldsymbol{\Phi}^{\dagger}\boldsymbol{v}_1
    \label{eq:amplitude}
\end{equation}

\section{ACTIVE FLOW CONTROL BASED ON DRL}\label{sec4}
Artificial neural network (ANN), as a basement of RL, is briefly introduced as follows.
As  a mathematical model for biological neural network, ANN is a system with many artificial neurons connecting by some topology.
An artificial neuron can compute the output with the inputs passed from connected neurons with adaptive weights and bias, which are updated by an algorithm like stochastic gradient descent to minimize a cost function on a training set~\cite{goodfellow2016deep}.
ANN has excellent nonlinear mapping ability and performs well on making a nonlinear link between the inputs and outputs~\cite{lecun2015deep}.
A multilayer feed-forward network with enough neurons in hidden layers can approximate any continuous function with arbitrary accuracy~\cite{hornik1991approximation}.

The DRL model, as an intelligent system with learning ability, can be applied in active flow control.
As one of the most important part of DRL, the environment in this study is the direct numerical simulation (DNS) of the flow field, as described in Section~\ref{sec2}.
With three basic elements related to the study, the DRL can be applied on the active control of the flow pass a circular cylinder.
They are: the state $s_t$ (here, an array of point measurements of velocity obtained from the simulation), the action $a_t$ (here, the active control of the jets, imposed on the simulation by the learning
agent) and the reward $r_t$ (here, objective function computed from the flow field information related to the stability of the vortex alley, which will be described in detail later).
After these three elements determined, the learning agent can interact with the environment by: choosing the action $a_t$ according the current state $s_t$, performing the action at the state in the environment and computing the reward $r_t$ and the next state $s_{t+1}$.
With numerous interactions with the environment, the learning agent is trained to choose action from the current state that yields the highest possible reward.
In this study the DRL agent can learn the control experience of flow past a circular cylinder from interaction with the DNS-based environment, which will guide the adjustment of mass flow of jets.

The reinforcement learning algorithm used in this paper is proximal policy optimization (PPO)~\cite{schulman2017proximal}.
Different from the value-based learning methods, e.g.,~deep Q-learning (DQN)~\cite{mnih2015human}, which is restricted to discrete and low-dimensional state and action space, PPO, as a policy-based learning method, can handle the high-dimensional action problems with the continuity of state and action space.
Another advantage of PPO is that it is less complex mathematically and requires little to no metaparameter tuning.
Like many other policy-based methods, PPO uses an actor-critic structure~\cite{konda2000actor}.
Two ANNs are used in the actor-critic structure, with the network parameters $\Theta^{\mu}$ and $\Theta^Q$, respectively.
The actor network approximates to the policy function $a=\mu(s|\Theta^\mu)$, which learns a policy that yields the highest possible reward $r$ so that a definite action from the current state can be given.
The critic network approximates the action-state value function $Q(s,a|\Theta^Q)$ to criticize the given action according to the action $a_t$ in state $s_t$ and following policy $\mu$.

During the training process, the action is obtained from the state $s_t$ for each step with the actor network as $a_t=\mu(s_t)$.
Then the agent interacts with the environment according to the action $a_t$ and the state $s_t$, feed back with the next state $s_{t+1}$ and the reward $r_t$.
The parameters of the actor network, $\Theta^{\mu}$, and the critic network, $\Theta^Q$, are updated according to the reward $r_t$ in each step.
The PPO method is episode-based, which means that the interactions between the agent and the environment are broken into a number of training interaction sequences~\cite{sutton2018reinforcement}.
As the PPO algorithm has already been used in a variety of fluid mechanics works, the reader interested in more details on the PPO algorithm itself is invited to refer to its previous application on AFC~\cite{RN3}.

The framework of DRL network used in this paper is built based on Tensorforce~\cite{tensorforce}, an open-source DRL library builds upon Tensorflow. 
The learning is performed in parallel with the multi-agent technique developed by Rabault~\cite{rabault2019accelerating}.
The structure and the hyperparameters of the PPO model are determined by referring the previous work on this topic of Rabault and Kuhnle~\cite{rabault2019accelerating}.
The ANNs are composed of two dense layers of 512 fully connected neurons, plus the input layer and the output layer.
For each step PPO agent interacting  with the environment, 50 numerical time steps  
are computed by the direct numerical simulation, corresponding to approximately $7.5\%$ of the vortex shedding period.
Therefore, the action time step $T=50dt,~(dt=5\times10^{-3}$), which means the PPO agent is allowed to interact with the simulation and update its control only each 50 time steps.
The total steps of an episode is chosen as 4000, with the episode duration $T_{total}=20.0$, which spans approximately 6.5 vortex shedding periods.
For each episode, 80 actions are determined by the network and performed by the agent.
The hyperparameters are chosen as: learning rate is $1\times10^{-3}$ and likelihood-ratio-clipping is 0.2.

The reward $r_t$, as the feedback from the simulation, varies by the determination of researchers and affects the final training result and learnt policy of the DRL model.
In this paper, to utilize the complete flow field information, DMDc is performed on the snapshots obtained from simulation in the training process to achieve a reward, to instruct the agent in learning a more global control policy.
For reducing the computation of DMDc, the flow field snapshot matrices composed with 30 columns of pressure measurements are used in this study.
The control snapshot matrices composed with 30 columns of actions of the agent. 
Therefore, the snapshot matrices $\boldsymbol{X}_0$, $\boldsymbol{X}_1$ and $\boldsymbol{\Gamma}$ in \ref{eq:snapshots} can be written as:\begin{equation}
    \boldsymbol{X}_0=\{\boldsymbol{p}_{t-28},\boldsymbol{p}_{t-27},...,\boldsymbol{p}_{t-1},\boldsymbol{p}_t\},
\end{equation}
\begin{equation}
    \boldsymbol{X}_1=\{\boldsymbol{p}_{t-27},\boldsymbol{p}_{t-26},...,\boldsymbol{p}_{t},\boldsymbol{p}_{t+1}\},
\end{equation}
\begin{equation}
    \boldsymbol{\Gamma}=\{\boldsymbol{a}_{t-28},\boldsymbol{a}_{t-27},...,\boldsymbol{a}_{t-1},\boldsymbol{a}_t\},
\end{equation}
where $p_t$ and $p_{t+1}$ refer to the pressure measurements of the state $s_t$ and the next state $s_{t+1}$. $p_{t-i}$ and $a_{t-i}$ refer to the pressure measurement and the action at $i$ time intervals before the current state, respectively.
The time interval of each adjacent measurement are set to 50 numerical time steps, equal to the action time step of the agent, so that the matrices $\boldsymbol{X}_0$, $\boldsymbol{X}_1$ and $\Gamma$ can be updated each time the agent determines the action and interacts with the simulation.
The 30 snapshots span approximately 4 vortex shedding periods.

To determine the expression of the reward, the modes of the baseline flow (with no actuation, i.e. $Q_1^*=Q_2^*=0$) is first studied by performing DMD on the snapshots of the complete flow field. 
The dynamic modes of the flow are $\phi_i,~i=0,2,...,28$, with corresponding eigenvalues $\mu_i,~i=0,2,...,28$ and modal amplitudes $\alpha_i,~i=0,2,...,28$ computed according to Eq.~\ref{eq:amplitude}.
The eigenvalues with non-negative imaginary part and the spectrum of the dynamic modes are illustrated in Fig.~\ref{fig:spectrum}.
The zeroth mode is the time-averaged flow  and has a reflection symmetry about the centerline $y=0$.
The first mode, as the main mode, corresponds to the part of the flow field that oscillates with the fundamental vortex shedding frequency.
This mode has the same spatiotemporal symmetry and the same streamwise spacing between the vortices as the full nonlinear cylinder flow.
The second mode and the third mode are due to the interaction of the first mode with themselves.
The second harmonic oscillates with twice the fundamental frequency, while three times the fundamental frequency for the third mode.

\begin{figure}[H]
\centering
\subfigure[]{\includegraphics[width=0.385\textwidth]{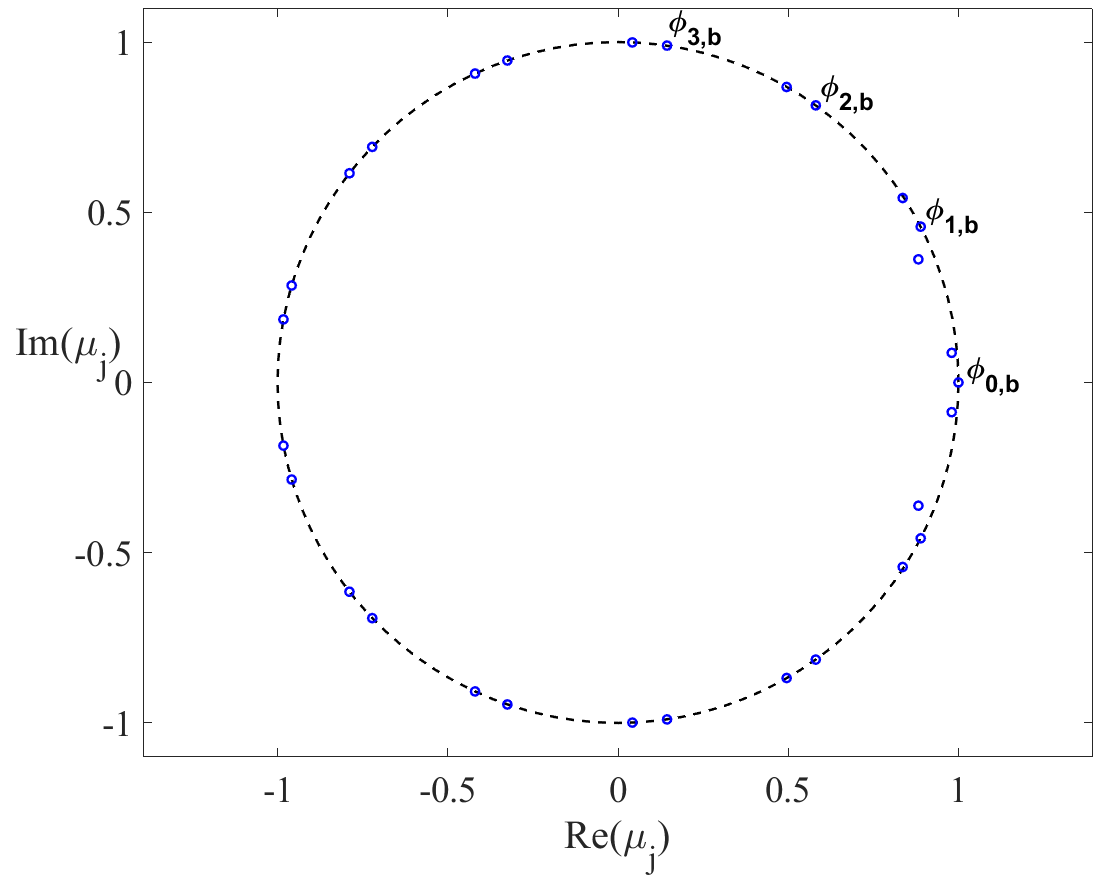}\label{fig:eigenvalues0}}
\subfigure[]{\includegraphics[width=0.57\textwidth]{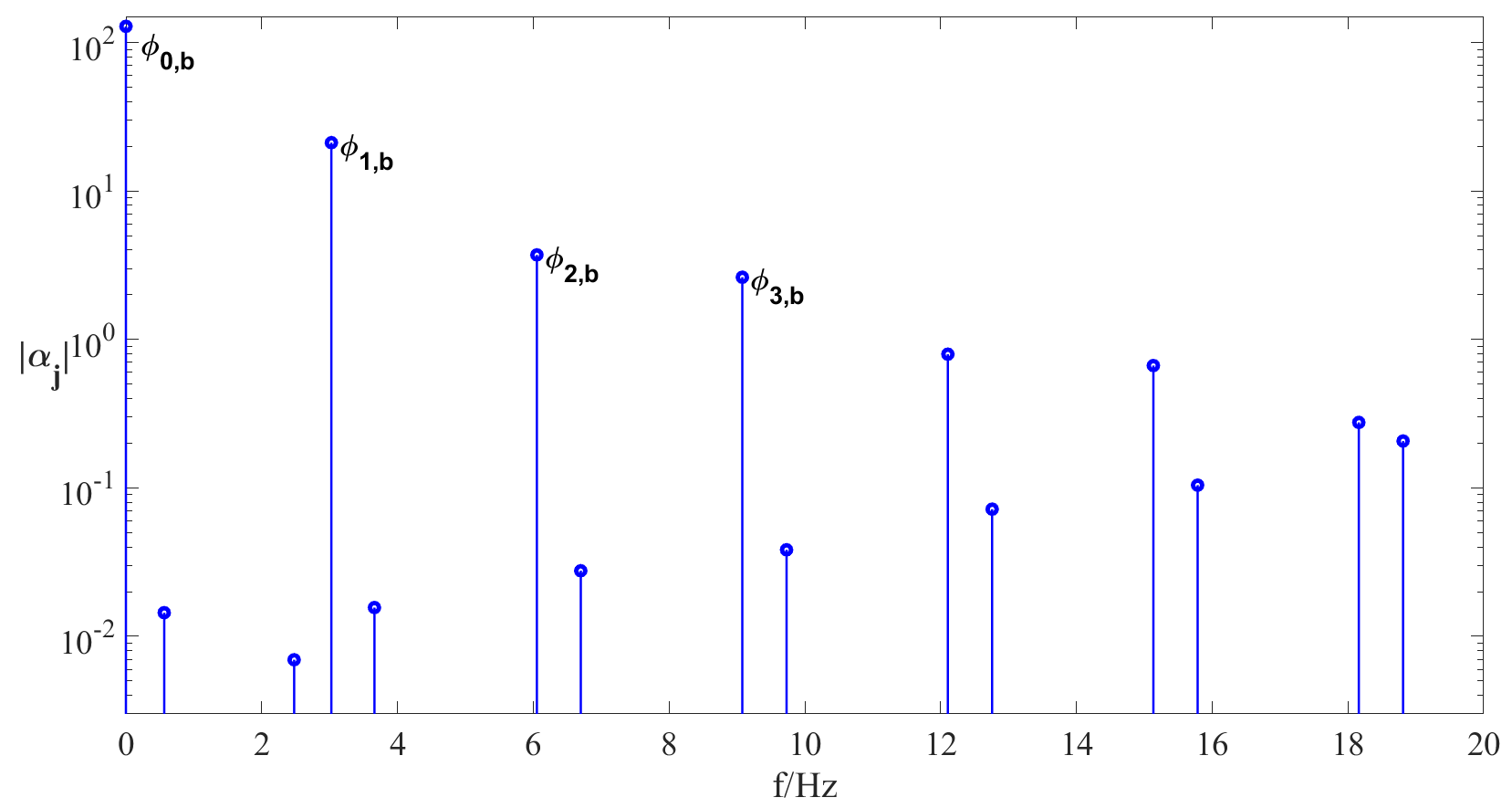}\label{fig:amplitude0}}
\caption{The eigenvalues (a), and the spectrum (b), of the dynamic modes of the baseline flow.
The eigenvalues and the amplitudes of the zeroth to the third modes are shown with $\phi_{i,b}$ representing the $i$th mode of the baseline flow.}
\label{fig:spectrum}
\end{figure}

The zeroth mode an energetic mode corresponding to the time-averaged flow. 
The other modes are corresponding to the unsteady flow.
In addition, the amplitudes of the other modes can also present the irregularity of the transient flow, e.g., for the baseline flow without actuation, the amplitudes of the other modes are smaller than those of the flow with stochastic mass flow rate of the jets.
Therefore, to obtain a relatively stable flow over a circular cylinder, the global kinetic energy of the flow is expected to decrease.
We hope that controlled flow can evolved into a regular flow like periodic flow, so the amplitudes of the other modes is expected to decrease either.
With some test on the flow with actuation, we find that the amplitudes of the other modes vary greatly according to the irregularity mass flow rate $Q_i^*$, which make it difficult for the agent to learnt the policy.
Thus, the amplitude of the zeroth mode is considered more in the reward to make sure the learning of the agent is stable and robust.
The expression of the reward $r_t$ are determined as follows: \begin{equation}
    r_t=0.198-0.1|\alpha_0|-0.0005\sum_{i=1}^{28}|\alpha_i|.
    \label{eq:reward}
\end{equation}
The weights in Eq.~\ref{eq:reward} are chosen to make terms of the first mode and the other modes are close in magnitude, and the bias are chosen to make the reward close to zero.

The training of the PPO model in this paper is performed on Intel~Xeon~CPU~E7-8890~v4 (48 cores @2.20GHz).
With the multi-agent technique, 80 agents perform and interact with the environment at the same time in parallel.
400 epochs of training \underline{process }can be finished in 5 periods with 80 epochs done in each period.
The time cost for one period of 80 epochs is about 40 minutes.

\section{RESULTS AND DISCUSSIONS}\label{sec5}

\begin{figure}[H]
\includegraphics[width=0.9\textwidth]{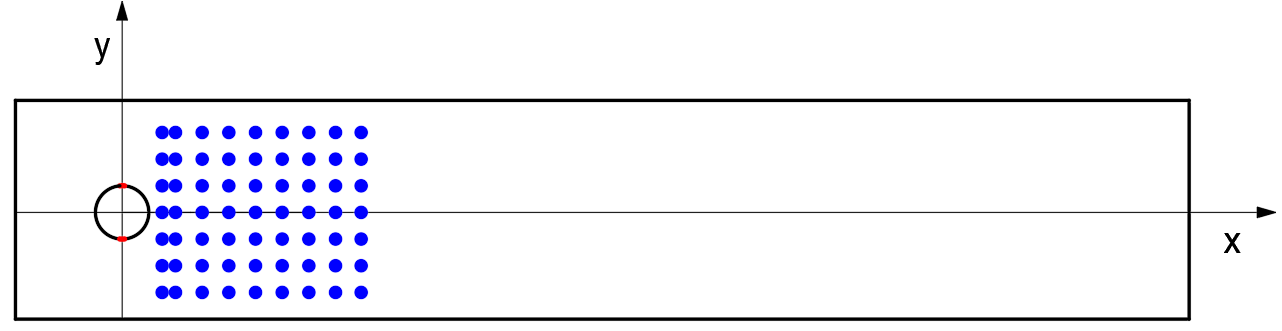}
\caption{\label{fig:probe} Positions of the probes (marked by blue dots) in the two-dimensional domain of the flow over a circular cylinder.}
\end{figure}

For the actual application of this AFC method in experiments, 63 probes are set in the wake of the cylinder to measure the flow physical quantities e.g. pressure and velocity, shown in Fig.~\ref{fig:probe}.
With the reward in Eq.~\ref{eq:reward} computed from the DMDc of the snapshots of the pressure measurements, the DRL-based AFC method is applied with a training process of 400 epochs.

\begin{figure}[H]
\centering
\subfigure[]{\includegraphics[width=0.8\textwidth]{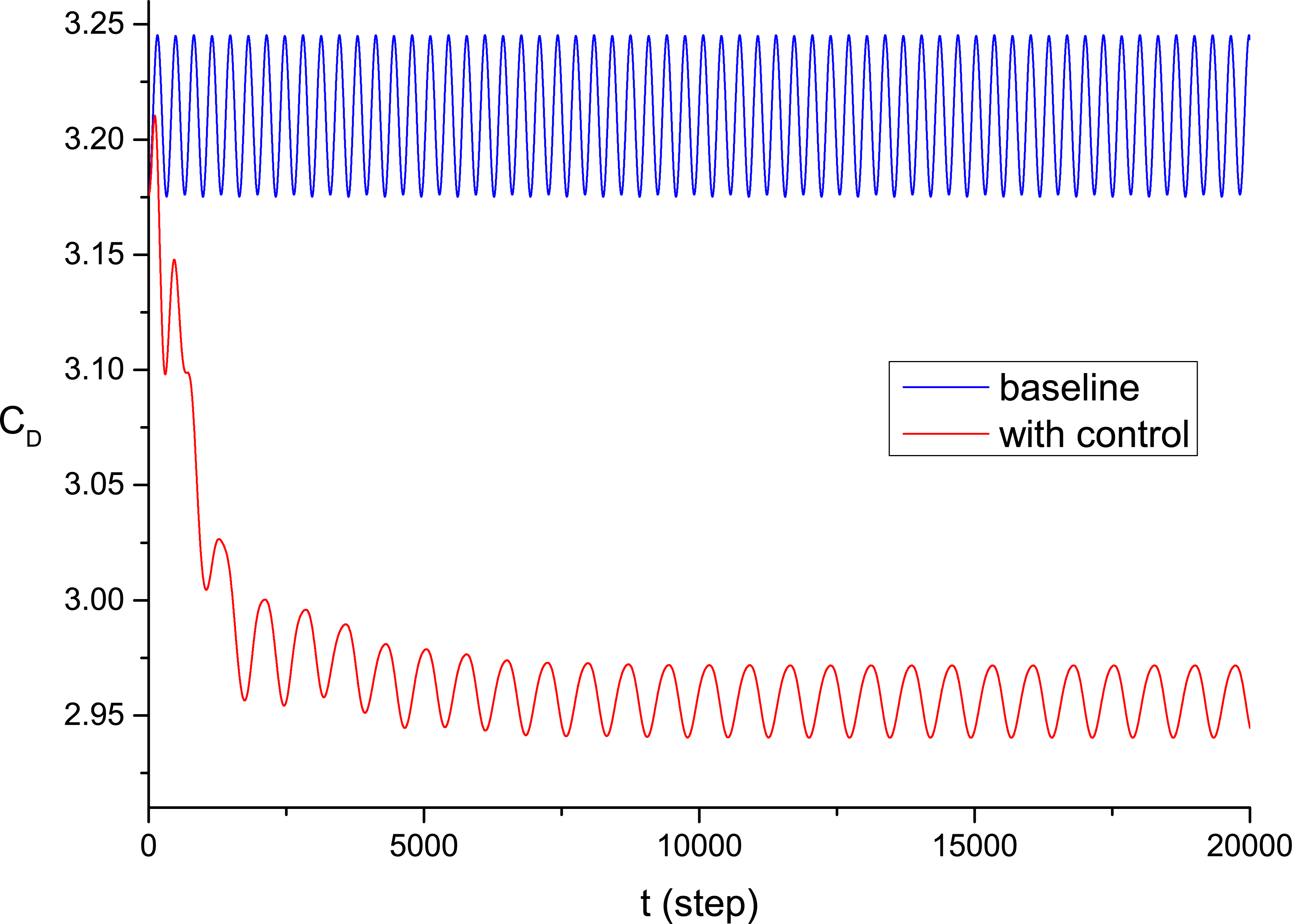}\label{fig:Cd}}
\subfigure[]{\includegraphics[width=0.8\textwidth]{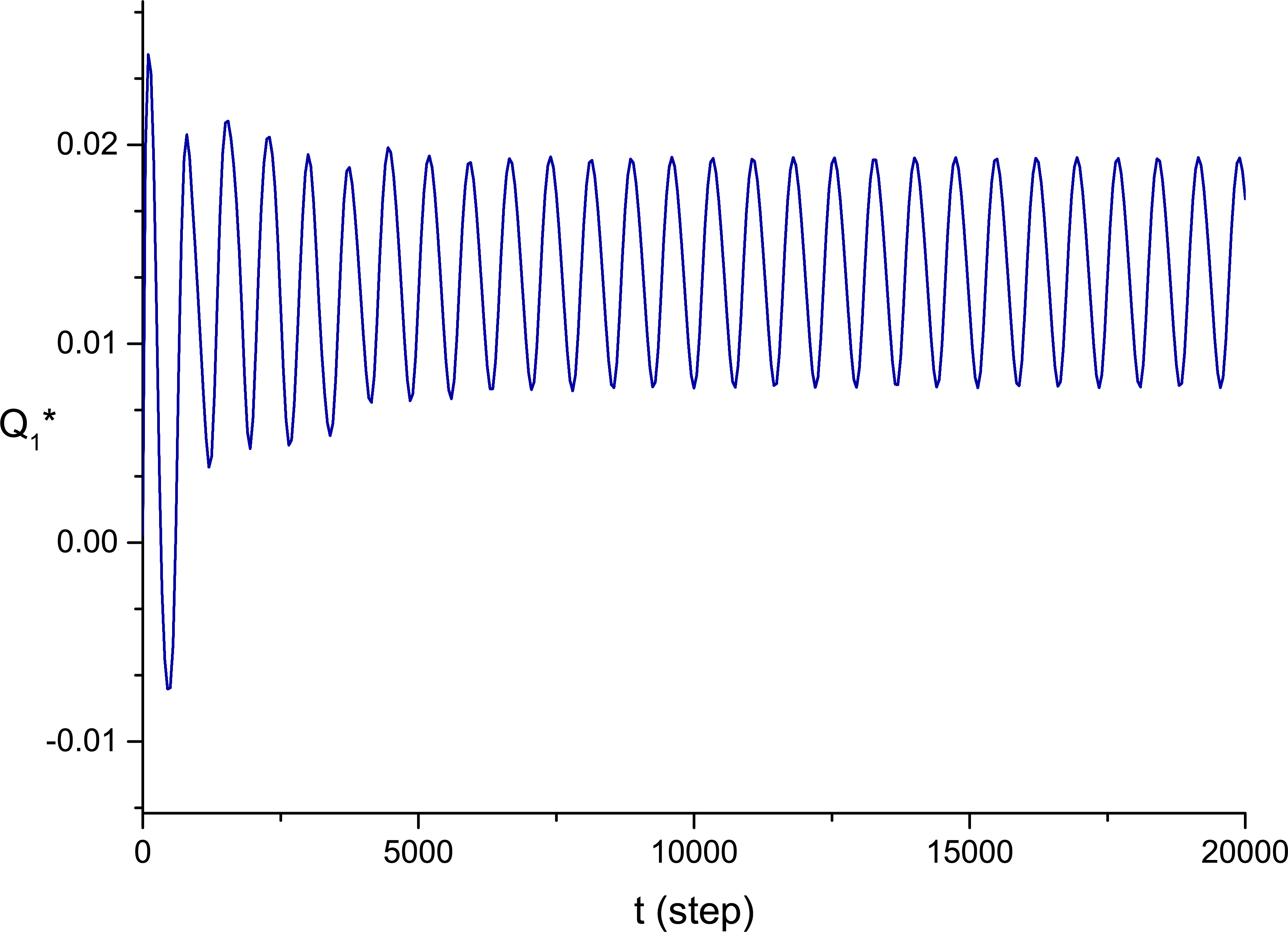}\label{fig:Q}}
\caption{ Time series for the coefficients of AFC. (a) Time-resolved value of the drag coefficient $C_d$ of the baseline flow (with no control, i.e. $Q_1^*=Q_2^*=0$) and the flow with control. (b) Time-resolved value of the normalized mass flow rate of one jet $Q_1^*$.}
\end{figure}

After the agent is trained, it is tested in the simulation-based environment of the AFC.
The mass flow injected by the jets is guided according to the learnt policy.
The drag coefficient $C_D$ curves of the baseline and controlled flow depending on the numerical time steps are plotted in Fig.~\ref{fig:Cd}.
The corresponding normalized mass flow rate of the jets $Q_1^*$ are shown in Fig.~\ref{fig:Q}.
With the robust policy learnt through DRL, the agent can stabilize the flow in about 2500 time steps.
As can be seen, the model with DMD-based reward is able to reduce the drag of the flow over a cylinder by applying AFC following training through DRL.
For the relatively stable stage in the control process (after 10000 steps), the mean value of $C_d$ is reduced from $3.21$ (baseline) to $2.958$ (with control), which represents a drag reduction of approximately $8\%$.
In addition to this reduction in drag, the fluctuations of the drag coefficient are reduced to approximately $0.0315$ by the active control, i.e. a decrease of roughly $55\%$ compared with the value of $0.0701$ for the baseline. 
Similarly, fluctuations in lift are reduced, from $2.12$ (baseline) to $0.807$ (controlled), with a factor of approximately $0.38$.
With FFT, the Strouhal number of the mass flow of the jets of the relatively stable stage, $St_{jet}$, are computed.
Similar to the $St$ of the controlled flow,  $St_{jet}=0.2686$, is approximately $0.89$ times the $St$ of the flow without actuation.
With the actuation evolving into an approximately periodic signal, the characteristic frequency of the system is modified.
The frequency of the vortex shedding of the controlled case is approximately $11\%$ lower than the baseline case.

\begin{figure}[H]
\centering
\begin{tabular}[c]{cc}
  \begin{tabular}[c]{c}
    \subfigure[~baseline]{\includegraphics[width=0.7\textwidth]{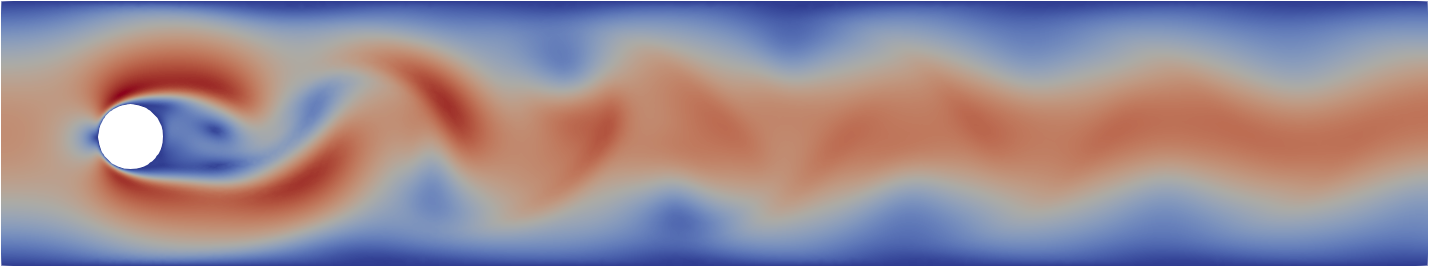}}\\
    \subfigure[~with control]{\includegraphics[width=0.7\textwidth]{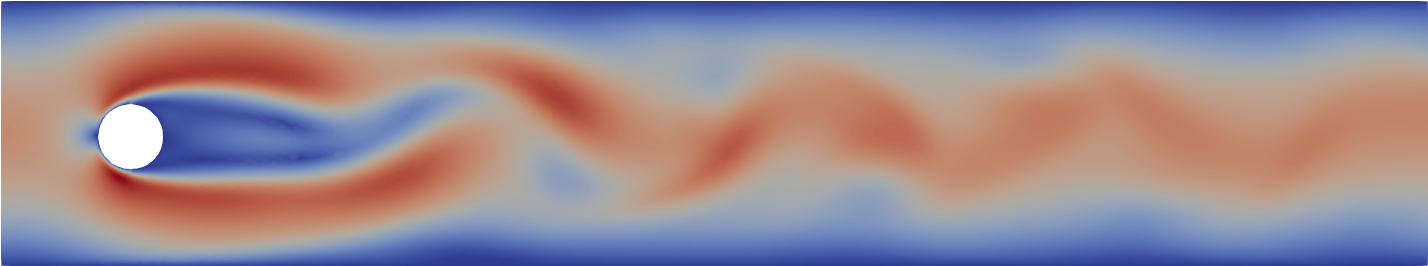}}
  \end{tabular}
    &
    \begin{tabular}[c]{c}
        \includegraphics[width=0.16\textwidth]{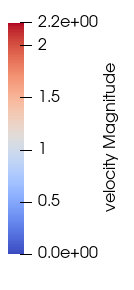}
    \end{tabular}
  \end{tabular}
\caption{\label{fig:Contours} The comparison of velocity magnitude contours of the 2D flow over a circular cylinder without actuation (a), and with active control (b), at $T=100$.}
\end{figure}

The velocity magnitude contours of the baseline flow and the flow with control are illustrated in Fig.~\ref{fig:Contours}.
With the control policy, the jets modify the flow configuration and stabilize the vortex alley.
In the case with active control, the low-velocity region of the wake behind the cylinder is extended and larger than in the baseline case.
The velocity ﬂuctuations induced by the vortices are globally less strong, and less active close to the upper and lower walls.
The wake of the vortices is increased in width and reduced in velocity magnitude.
For numerical comparison, the recirculation area, defined as the region in the downstream neighbourhood of the cylinder where the horizontal component of the velocity is negative, are measured in both cases.
The average recirculation area of the controlled flow is $0.0223$, with a relative improvement of $109\%$ compared with the average recirculation area of $0.01067$ for the baseline flow.
With the active flow control, dramatically increase is achieved in the recirculation area, which shows the efﬁciency of the learnt control policy at reducing the effect of vortex shedding.

\begin{figure}[H]
\centering
\subfigure[]{\includegraphics[width=0.385\textwidth]{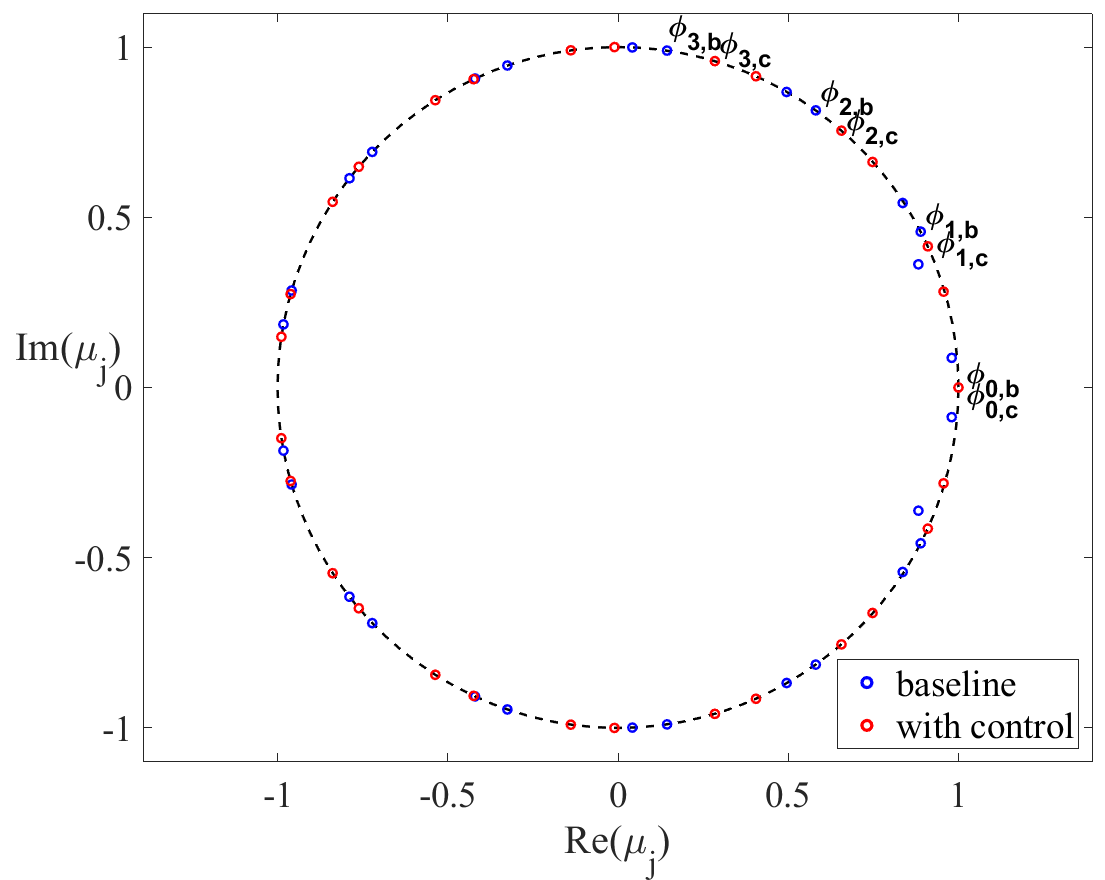}\label{fig:eigenvalues}}
\subfigure[]{\includegraphics[width=0.57\textwidth]{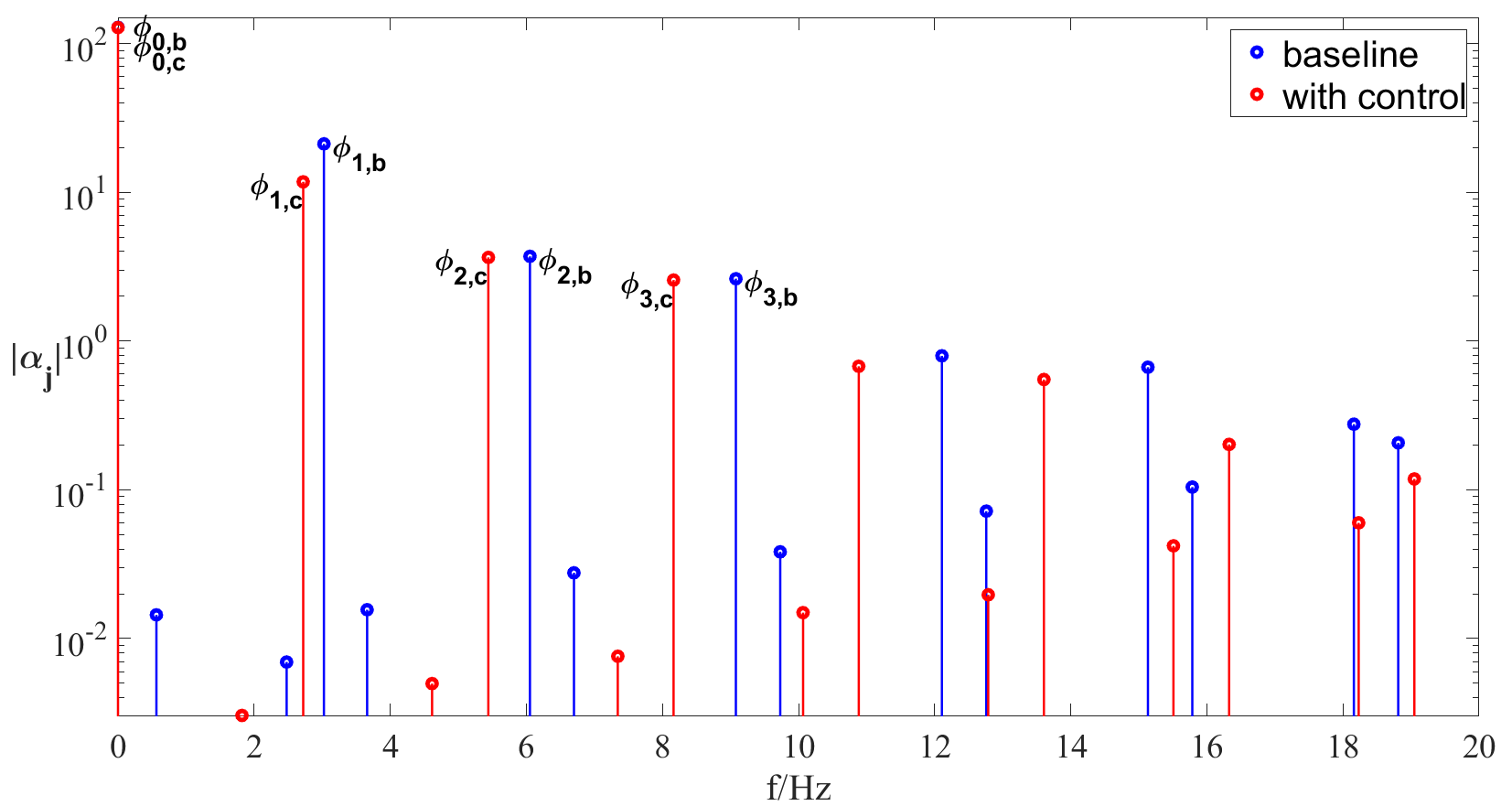}\label{fig:amplitude}}
\caption{\label{fig:spectrum}The eigenvalues (a), and the spectrum (b), of the dynamic modes of the baseline flow and the controlled flow.
The eigenvalues and the amplitudes of the zeroth to the third modes are mark in the figure.
$\phi_i,b$ represents the $i$th mode of the baseline flow while $\phi_i,c$ represents the $i$th mode of the controlled flow.}
\end{figure}

The DMD are performed on the complete flow field snapshots of the baseline and controlled cases for more detailed analysis.
Snapshot matrices composed of 50 columns of measurement on the total grid points, with a non-dimensional time interval of $0.25$ are used for each case.
With a truncation of 29, the number of the dynamic modes is equal to the modes used to computed the reward.
The imaginary verses real part of the eigenvalues of the baseline flow and the controlled flow, $\mu_{j,b}$ and $\mu_{j,c}$, respectively, is plotted in Fig.~\ref{fig:eigenvalues}.
With the computed amplitudes $\alpha_{j,b}$ (baseline) and $\alpha_{j,c}$ (with control), the spectrum is shown in Fig.~\ref{fig:amplitude}.
As can be seen from the spectrum, in the controlled case, the frequency of the first mode (fundamental frequency) is decreased when compared with the baseline case, as well as the frequencies of the second and third mode.
With the AFC, the amplitude of the zeroth mode is sightly reduced.
The amplitude of the first mode (main mode) is greatly reduced, with an approximately $45\%$ reduction
The amplitudes of the second and the third mode both decrease, while the other modes not marked in the Fig.~\ref{fig:amplitude} also have large reductions on amplitudes.
The results show that though the amplitude of the zeroth mode is considered more while the amplitudes of the first to fourteenth are considered less in the reward, based on the modified reward, the agent still can learn a policy of AFC to reduce the amplitudes of first to fourteenth modes greatly.
The large reductions on the amplitudes of the modes also show the AFC learnt by the method of this paper is effective in stabilizing the flow over a circular cylinder.

\begin{figure}[H]
\centering
\begin{tabular}[c]{cc}
  \begin{tabular}[c]{c}
    \subfigure[~$0^{th}$ mode, baseline]{\includegraphics[width=0.73\textwidth]{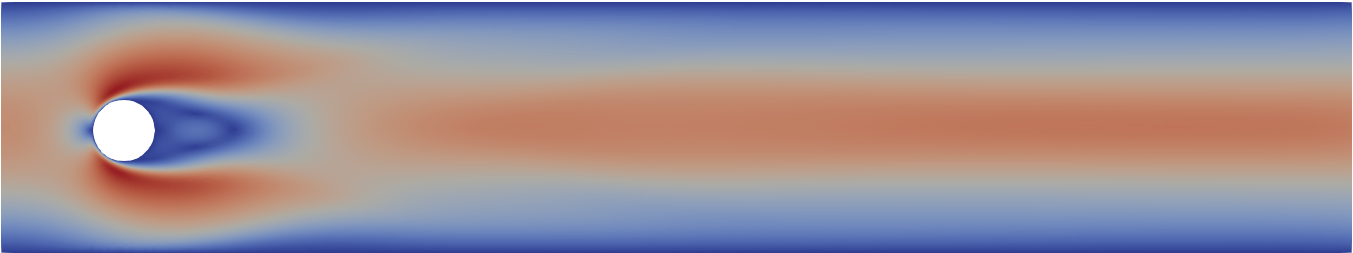}\label{fig:0th_b}}\\
    \subfigure[~$0^{th}$ mode, with control]{\includegraphics[width=0.73\textwidth]{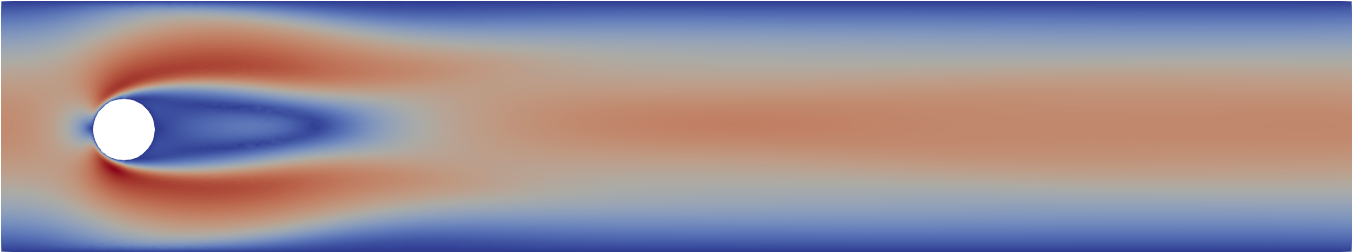}\label{fig:0th_c}}
  \end{tabular}
    
    \begin{tabular}[c]{c}
        \includegraphics[width=0.14\textwidth]{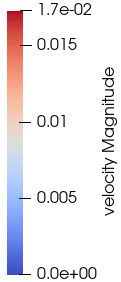}
    \end{tabular}
    \\
  \begin{tabular}[c]{c}
     \subfigure[~$1^{st}$ mode, baseline]{\includegraphics[width=0.73\textwidth]{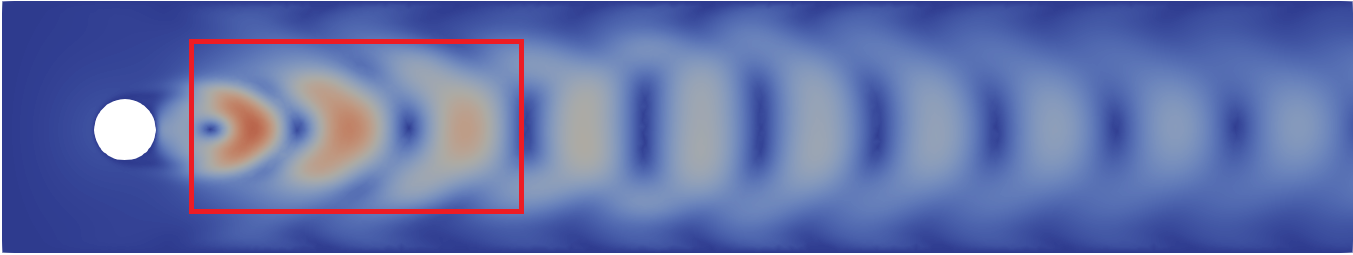}\label{fig:1st_b}}\\
    \subfigure[~$1^{st}$ mode,with control]{\includegraphics[width=0.73\textwidth]{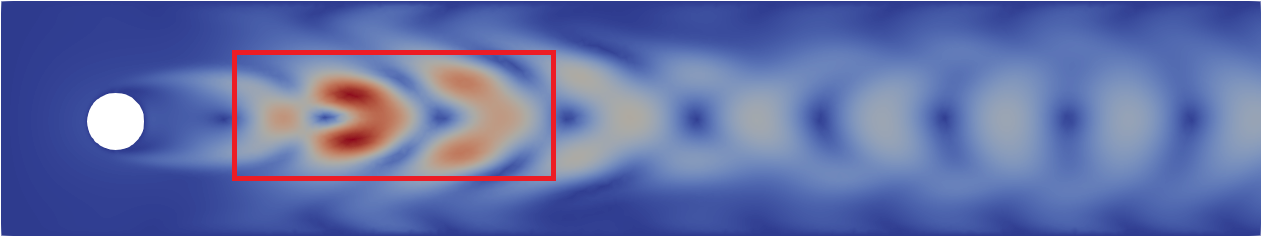}\label{fig:1st_c}}
    \end{tabular}
    
    \begin{tabular}[c]{c}
        \includegraphics[width=0.14\textwidth]{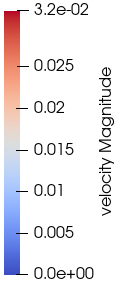}
    \end{tabular}
    \\
    \begin{tabular}[c]{c}
    \subfigure[~$2^{nd}$ mode, baseline]{\includegraphics[width=0.73\textwidth]{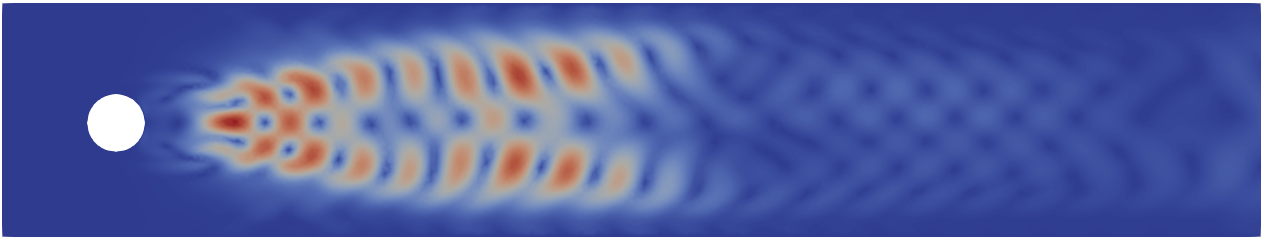}}\\
    \subfigure[~$2^{nd}$ mode, with control]{\includegraphics[width=0.73\textwidth]{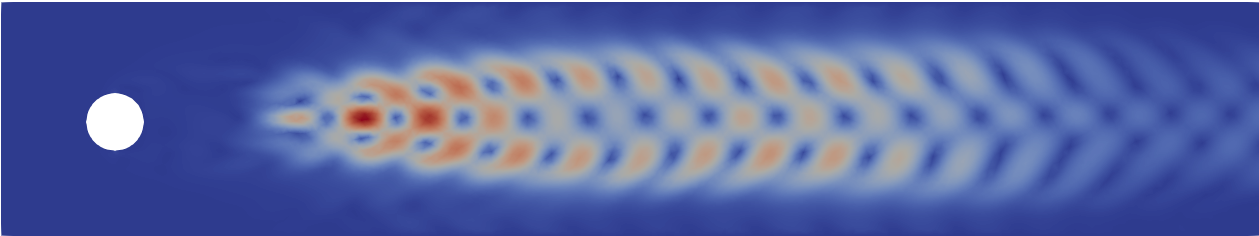}}
  \end{tabular}
    
    \begin{tabular}[c]{c}
        \includegraphics[width=0.14\textwidth]{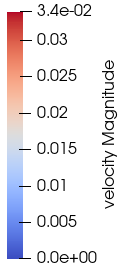}
    \end{tabular}
    \\
  \begin{tabular}[c]{c}
     \subfigure[~$3^{rd}$ mode, baseline]{\includegraphics[width=0.73\textwidth]{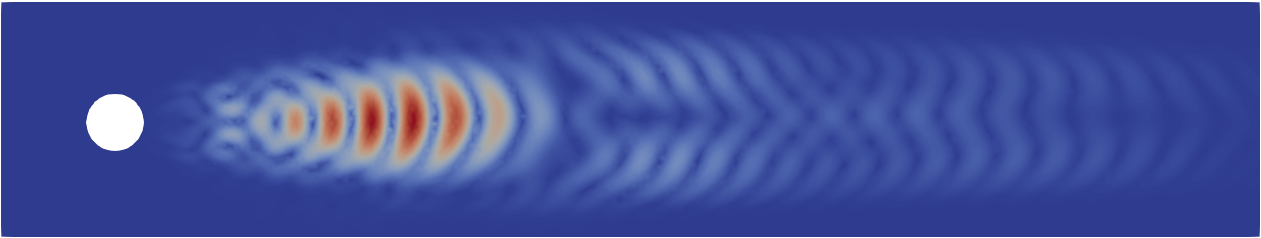}}\\
    \subfigure[~$3^{rd}$ mode, with control]{\includegraphics[width=0.73\textwidth]{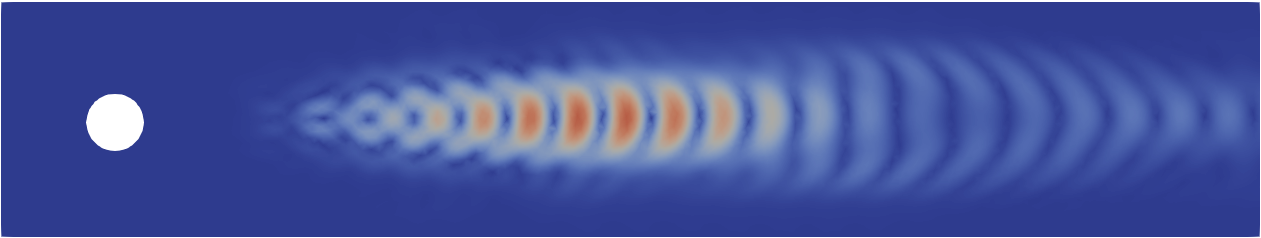}}
    \end{tabular}
    
    \begin{tabular}[c]{c}
        \includegraphics[width=0.14\textwidth]{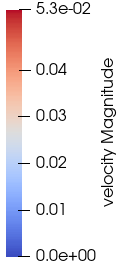}
    \end{tabular}
  \end{tabular}
\caption{\label{fig:dmdmodes} The comparison of velocity magnitude contours of the zeroth to the third modes of the baseline flow and the controlled flow reconstructed through DMD on respective velocity snapshots.}
\end{figure}

The velocity magnitude contours of the zeroth to the third dynamic modes of the flow without actuation and with AFC reconstructed from the DMD on velocity measurement are plotted in Fig.~\ref{fig:dmdmodes}.
The zeroth modes (Fig.~\ref{fig:0th_b},~\ref{fig:0th_c} ) show the similar modification as in the flow field velocity magnitude contours (Fig.~\ref{fig:Contours}), like extended low-velocity region behind the cylinder, larger wake width and reduced velocity magnitude of the wake in the controlled case.
For the first to the third mode, with the AFC, the space of the fluctuation of the mode increases, consistent with the change of the frequencies in FIG.\ref{fig:spectrum}.
The starting position of the modes (i.e. the nearest horizontal position with nonzero measurement in the downstream of the cylinder) moves aft along the positive x-axis, as well as the region of larger velocity magnitude.
The width of the region of the modes with non-zero velocity is reduced by the AFC, 
which means the area affected by the flow pass the cylinder is reduced.
For the first mode, the velocity amplitude at the approximate position of $x=5$ (marked with red boxes in Fig.~\ref{fig:1st_b},~\ref{fig:1st_c}) is increased, which represents the main mode actuated by the control jets.

\begin{figure}[H]
    \centering
    \includegraphics[width=0.9\textwidth]{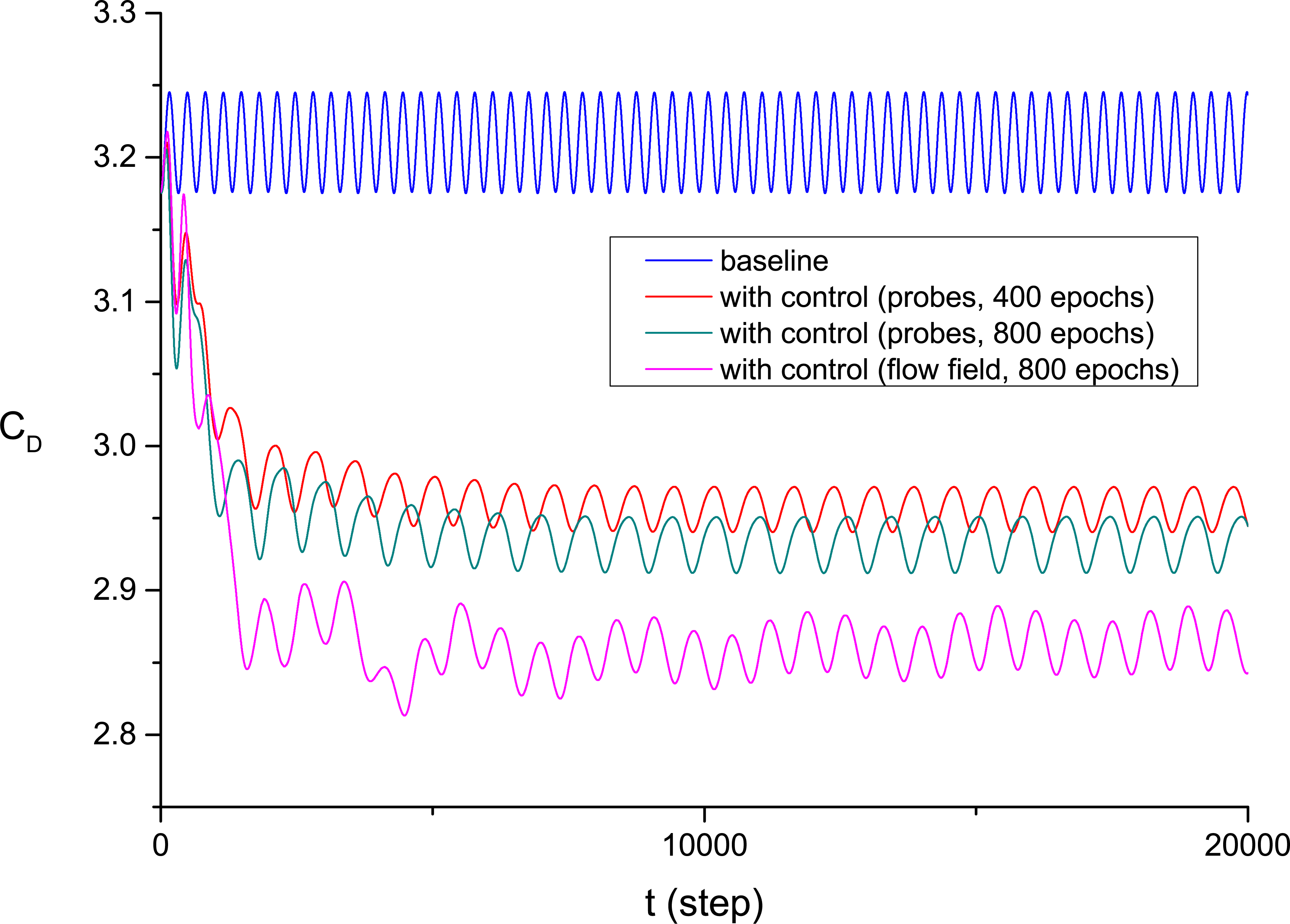}
    \caption{Time-resolved value of the drag coefficient $C_d$ of the flow with different configurations.
    (Blue curve: baseline flow; red curve: controlled flow with the probe-based reward trained by 400 epochs; cyan curve: controlled flow with the probe-based reward trained by 800 epochs; purple curve: controlled flow with the field-based reward trained by 800 epochs.)}
    \label{fig:Cds}
\end{figure}

Furthermore, training with different epochs are tested in this paper.
For the case of reward based on the probe measurement described above, the drag coefficient $C_d$ curves of training processes of epochs of 400 and 800 are shown in Fig.~\ref{fig:Cds}.
The agent achieves a better control when the training epochs is doubled.
The reduction of the mean value of $C_d$ at the relatively stable stage is approximately $8.6\%$.
The mean value of the recirculation area is $0.0239$, with an improvement of $124\%$, compared with $109\%$ of the 400-epoch case.
In addition, a study with the reward based on the DMDc of the full pressure information of the complete flow field is also done.
In the training process of the DRL agent, DMDc is performed on the spanshots composed of 12837 pressure values of total grid points.
For the more high-dimensional information compressed by DMDc, the agent needs more steps to learn from the interaction with the environment.
It takes 800 epochs for the reward of the DRL model based on information of the complete flow field decreases to a low value.
The drag coefficient $C_D$ curves of the reward based on information of probes and the complete flow field are plotted in Fig.~\ref{fig:Cds}, compared with the case of probe measurement (training of 800 epochs).
The reward based on the complete flow field brings a great improvement on the reduction of the drag.
The reduction of the mean value of $C_d$ at the relatively stable stage is approximately $10.8\%$, which is a large improvement by comparison with $8.6\%$ of the probe-based case.
The results shows the good robustness of this method on different training periods and measurement resolution.
Moreover, there are approaches to improve the AFC results of this method further, by extending the learning process of the agent and using more complete information of the flow field.

\section{CONCLUSIONS}\label{sec6}
In this study, the framework initially presented in the work of Rabault et al.~\cite{RN3} is extended by using a reward of more flow information based on DMDc in the DRL/PPO model.
The DRL agent can learning the AFC policy for synthetic jets on a cylinder, and control the conﬁguration of the 2D K\'arm\'an vortex street, through the optimization of a reward function modified in this paper.
The improved reward function are composed with the amplitudes of the dynamic modes of the flow, which represents the main dynamic characteristics extracted from the detailed flow field information.
With the AFC, a drag reduction of approximately $8\%$ is achieved in the case of reward computed from the DMDc on probe measurements, while a drag reduction of up to approximately $10.8\%$ is achieved in the case of reward based on the complete flow field.
An improvement of $109\%$ in the recirculation area is observed with the learning process of 400 epochs and an improvement of up to $124\%$ is obtained in the recirculation area with the learning process of 800 epochs.
The good results in drag reduction and improvement of recirculation area show the effectiveness of the control policy, as well as the framework based of the data-driven reward.
The AFC results of this method can be improved further by extending the learning process of the agent and using more complete information of the flow field.

From the further analysis through the DMD on the complete flow field information, the modifications of the modes for the flow with AFC can be observed.
The fundamental frequency of the controlled flow is decreased of approximately $11\%$.
With the active control the amplitudes of the modes of the flow are reduced with the active control, especially the main mode.
Larger space between the fluctuation of the mode and mode positions in further downstream of the cylinder are observed in the velocity magnitude contours reconstructed from the DMD-modes.
These results show that the modifications of the modes for the flow through control is beneficial to stabilize the vortex street, which also shows that the AFC policy learnt through with the improved reward is effective in stabilize the flow over a cylinder.

Since measuring probes are used in this study, the framework of the DRL model with DMD-based reward presented in this paper is able to be applied to experiments in the further study.
The configuration of the DRL model can be modified into the case with more steps of episode (larger episode duration) and less epochs of learning for the application in experiments.
As concluded from this study, with more probes positioned in the flow domain, more complete information of the field can be measured, which will results in an improvement in the control result.
Besides the experimental application, this method has many open prospects of research.
It can be applied to other AFC problems with more complex geometry like airfoils in the study of stall.
It also can be considered to be utilized in more complex simulations such as three-dimensional large eddy simulations.
In this case, the amplitudes of the dynamic modes are used in the computation of the reward function, while more information like the characteristics in the mode vectors can be considered in the reward function in the further study.

\bibliography{dmddrl.bib}

\end{document}